\documentclass[11pt]{article}
\usepackage{amssymb,amsmath}
\usepackage{graphicx}
\newcommand{\Ne}{\ensuremath{N_\mathrm{e}}}
\begin{document}
\begin{center}
\textbf{\Large On the angular distribution of extensive air showers}

\bigskip

M.Yu.\ Zotov, N.N. Kalmykov, G.V. Kulikov$^a$, V.P. Sulakov\\[1mm]

{\itshape
D.V. Skobeltsyn Institute of Nuclear Physics, Moscow State University, Russia
}

E-mail: $^a$\texttt{kulikov@eas.sinp.msu.ru}

\end{center}
{\narrower
Angular distributions of extensive air showers with different number of
charged particles in the range $2.5\times10^5$--$4\times10^7$ are derived
using the experimental data obtained with the EAS MSU array. Possible
approximations of the obtained distributions with different empiric
functions available in literature, are analysed. It is shown that the
exponential function provides the best approximation of the angular
distributions in the sense of the $\chi^2$ criterion.

}

PACS: 96.50.sd

Keywords: extensive air showers, angular distribution

\section*{Introduction}

The angular distribution of extensive air showers (EAS) reflects the
process of development and the following absorption of an air shower as
it traverses the atmosphere of the Earth. This is the reason why an
interest to the subject does not decrease in spite of the long-time
study.

The first results on the altitude development of EAS and their angular
distribution were obtained in the middle of the twentieth century, see,
e.g., a review by Greisen~[1]. The results related to air showers
detected at a fixed density of the particle flux.  The results of the
experiments for the lower third of the atmosphere witnessed about an
exponential absorption of showers with altitude.  As a rule, the
distribution of the zenith angle obtained from data on the altitude
development was approximated as $\sim\cos^n\theta$ with $n\approx8.3$ for
the sea level.

This result was confirmed more than once in numerous works in which
individual EAS characteristics, including those performed with the
KASCADE experiment, which was begun in 1996 especially for investigating
the problem of the knee in the primary cosmic ray energy spectrum at
$3\times10^{15}$ eV~[2].  However, it was claimed recently basing on the
results obtained with the Tien Shan experiment (atmosphere depth
690~g~cm$^{-2}$) that there is a considerable excess of air showers with 
$\cos\theta$ being equal to 0.6--0.7 for EAS with the size above $10^7$
particles~[3, 4]. The result was interpreted by the authors as an
indication of the increasing role of the so called ``long-flying
component'' in the longitudinal development of an EAS~[4].  Due to this,
it is appropriate to discuss the problem of the altitude development of
EAS and their angular distribution once again.

\section{Experimental Data}

In our paper, we present results of an analysis of angular distributions
of air showers in sufficiently narrow intervals in accordance with the
number of charged particles~\Ne{} in a wide range of~\Ne{} from
$2.5\times10^5$ up to $4\times10^7$.  Experimental data obtained with the
EAS MSU array from 1984 to 1990 were used for the analysis. A detailed
description of the array can be found in~[5].  The array covered an area
of $\sim0.5$~km$^2$ and included 77 detectors of density of the flux of
charged particles.  The values of~\Ne{} were determined with the help of
an empiric function of lateral distribution that provided the best fit
for the experimental data~[6].

Arrival directions of EAS, defined by the zenith and azimuthal angles
$\theta,~\phi$ respectively were determined using a system of fast
scintillation counters via measuring relative time delays of triggering
of these counters, located at different places of the observation plane.
Thirty-six counters with 5-cm thick scintillators having an area of
0.5~m$^2$ each were used for the measurements.  Eight counters were
located at the central unit of the array, six were placed at the distance
of about 60~m from it. Other 22 counters were placed uniformly over the
area covered by the array.

An arrival direction of each EAS was determined in two steps. The method
of least squares with a flat air shower front assumed was used as the
first approximation. The method of maximum likelihood was employed next.
The method took into consideration experimental data on the distribution
of particles over the depth of a shower disc and the curvature of a
shower front obtained with the EAS MSU array~[7].

Accuracy in determination of the zenith angle was of the order of
$3^\circ$ for the majority of EAS, improving for greater values of
$\theta$ and decreasing for nearly-vertical air showers. Methodical
errors increased the values of the zenith angles due to relatively lesser
delays of triggering of counters located closer to the shower axis
because of higher flux density of the incident particles and their
narrower temporal distribution.

\section{Main Results}

All showers registered with the probability greater than 0.95 were
selected for the analysis of angular distributions. The effective region
for selecting the registered EAS was determined taking into account
fluctuations in air shower development. It was determined by the
triggering system of EAS, the value of~\Ne{}, and the parameter of the
function of the lateral distribution of charged particles in each shower.
Air showers with $\theta>45^\circ$ were excluded from consideration. 
These EAS formed approximately 1\% of the whole data set. Air showers
with $\theta<6^\circ$ were also excluded from the analysis because of the
greatest errors in determination of their zenith angles. Radii of regions
of effective registration, intervals of~\Ne{} that include the knee in
the size spectrum of EAS at $\Ne\approx4\times10^5$, and the number of
air showers in each of the intervals are presented in the Table.

\bigskip
\noindent
{\bf Table}. Intervals of~\Ne{}, radii of regions of effective registration, the number of air showers in each interval, values of $P(\chi^2)$, and values of the absorption path~$\lambda$ of EAS.
\begin{center}
\begin{tabular}{|c|c|c|c|c|c|c|c|}
\hline
$\Delta\lg\Ne$ & 5.4--5.6 & 5.6--5.8 & 5.8--6.0 & 6.0--6.2 & 6.2--6.6 &
                 6.6--7.0 & 7.0--7.6 \\
\hline
$R_\mathrm{eff}$, m & 20  & 30 & 40 & 50 & 59 & 85 & 158 \\
\hline
Number of EAS  &12923  &13699  &10608  &7302  &5720  &1925  &1338  \\
\hline
$P(\chi^2)$, \%& 7   & 60 & 6  & 42 & 54 & 64 & 80  \\
\hline
$\lambda$, g~cm$^{-2}$&120  &115  &112  &110  &115  &120  &116  \\
\hline
\end{tabular}
\end{center}
\bigskip

Angular distributions for the listed intervals of~\Ne\ were obtained by
splitting the data for the zenith angle into $3^\circ$-wide intervals.
Figure~1 demonstrates angular distributions of air showers for three
intervals of~\Ne.

\begin{figure}[!ht]
\centerline{\includegraphics[height=0.87\textheight]{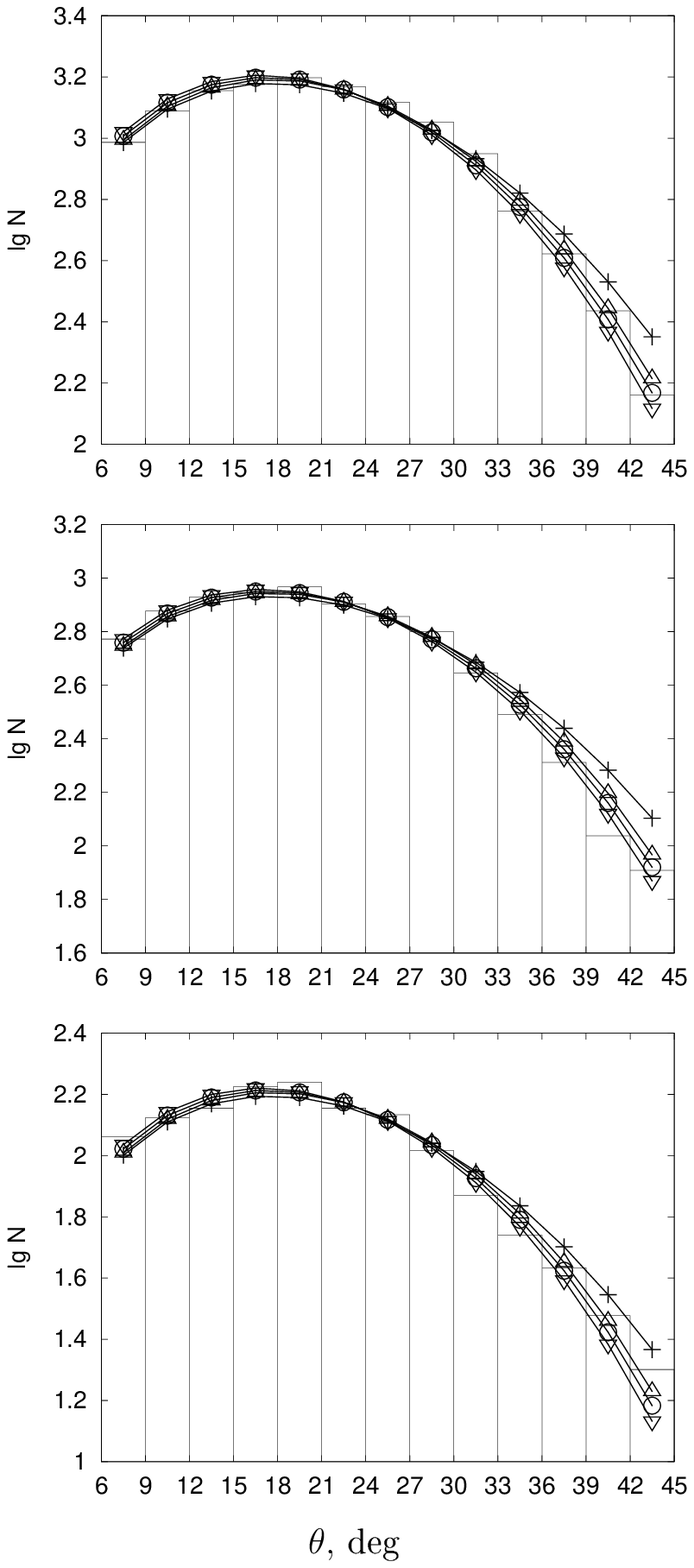}}

\caption{Angular distributions for $\Ne=5.4$--5.6, 6.0--6.2, 7.0--7.6
(from top to bottom). $N$ is the number of EAS in bins.
The curves show the behaviour of distribution (1) for
$\lambda=110$, 115, 120~g~cm$^{-2}$
($\bigtriangledown$, $\bigcirc$, $\bigtriangleup$ respectively) and
$\cos^9\theta$~($+$).}

\end{figure}

It is natural to compare the experimental angular distributions of EAS
with the exponential distribution expected \textit{a priori}:
\begin{equation}
	f(\theta) = A \exp \left[ -\frac{x_0}{\lambda} \left(\frac1{\cos\theta}
               - 1 \right)\right],
\end{equation}
as well as with its approximation given by $\cos^n\theta$. In the above
expression, $A$ is a normalization factor, $x_0$ is the vertical
atmosphere depth, $\lambda$ is the absorption path of air shower. In both
cases, an additional factor $\cos\theta$, which takes into account the
decrease of the effective area of the array with increasing angle of
inclination of an air shower, was used. The value of the absorption path
was found by minimizing $\chi^2$.

Results of the comparison of different distributions for the given
intervals of~\Ne\ are also shown in Figure~1.  It is clearly seen that
agreement between the experimental distributions and their approximations
is quite good, though for two of the intervals the probability
$P(\chi^2)$ is not high. Values of $P(\chi^2)$ together with the values
of the absorption path $\lambda$ are given in the Table.

The optimal values of the absorption path $\lambda$, which ensure a
minimum to $\chi^2$, were found with an error $\pm0.5$~g~cm$^{-2}$.  As
it follows from the Table, the absorption path of the number of air
showers remains almost constant in the considered range of~\Ne\ (the mean
value $\bar\lambda\approx115$~g~cm$^{-2}$ and the standard deviation
$\sigma\approx4$). Small values of $P(\chi^2)$ possibly relate to random
fluctuations.

For the sake of comparison, calculated angular distributions in the
usually accepted form $\sim\cos^n\theta$, where $n=x_0/\lambda$, are also
shown in Figure~1. In our case, $n=9$. As is clear from the figure, the
agreement of this approximation with the experimental data is worse,
especially for highly inclined air showers.

In the approximation that does not take into account fluctuations in the
development of an air shower, the absorption path of an EAS relates to
the absorption path $\Lambda$ of the number of particles in an air shower
as $\Lambda=\kappa\lambda$, where $\kappa$ is an exponent of the integral
size spectrum~[8].  For air showers with $\Ne\lesssim10^6$,
$\kappa\approx1.5$. For $\Ne\gtrsim10^6$, the value of $\kappa$ grows
approximately up to 2.0.  Therefore, under the assumption that $\lambda$
is almost constant in the considered range of EAS sizes, we come to the
conclusion that the development of an air shower slows down for great
values of~\Ne\ in spite of a change of the chemical composition of cosmic
rays toward heavier nuclei.

\section*{Conclusion}

Results comparable with the presented above were obtained at the sea
level with the KASCADE experiment during an investigation of EAS with the
number of particles~\Ne\ in the range from $3\times10^4$ to
$3\times10^6$, which includes the region of the knee in the size
spectrum~[2].  The absorption path of the number of particles $\Lambda$
in an air shower was found by the method of constant intensity to be
equal to 175~g~cm$^{-2}$ before the knee and to grow up to
194~g~cm$^{-2}$ above the knee.  Assuming another location of the knee in
the size spectrum, it was found in another work that also employed data
from the KASCADE experiment that $\Lambda=222\pm28$~g~cm$^{-2}$~[9].
Close results were obtained in the EAS-TOP experiment
($\Lambda=219\pm3$~g~cm$^{-2}$ at depth 820~g~cm$^{-2}$)~[10].

In conclusion, let us consider the results presented in~[4]. The authors
of that work reported an excess of EAS with zenith angles that correspond
to atmosphere depth greater than 1100~g~cm$^{-2}$ and have absorption
path $\lambda=585\pm45$~g~cm$^{-2}$ instead of
$\lambda=130\pm7$~g~cm$^{-2}$ observed for lesser angles. The EAS MSU and
KASCADE experiments do not confirm this result.

Thus, the analysis performed and the comparison of its results with the
results of the KASCADE experiment do not provide any ground for the
conclusion about an increasing role of the long-flying component with the
growth of the primary energy of cosmic rays.

\bigskip The work was performed with a financial support from the RFBR,
Grant No.08-02-00540.

\end{document}